\documentclass[runningheads]{llncs}

\usepackage{amsmath}
\usepackage{graphicx}
\usepackage{tikz-cd}
\usepackage{todonotes}
\usepackage{stmaryrd}
\usepackage{url}
\usepackage{mathtools}
\usepackage{ebproof}
\usepackage{upgreek}
\usepackage{newunicodechar}
\usepackage{orcidlink}
\usepackage{hyperref}
\newunicodechar{λ}{$\lambda$}
\newunicodechar{Λ}{$\Lambda$}
\newunicodechar{∀}{$\forall$}

\hypersetup{
  colorlinks = true,
  urlcolor   = blue,
  linkcolor  = blue,
  citecolor  = red
}

\newif\ifminted
\mintedtrue
\newif\ifincludeappendix
\includeappendixtrue

\ifminted

\usepackage[cachedir=_minted-cache,frozencache=true]{minted}
\setmintedinline{breaklines}
\newminted{Coq}{}
\newminted{haskell}{}
\newminted[appendixcode]{haskell}{
  linenos=true,
  escapeinside=@@, 
}

\AtBeginEnvironment{minted}{\dontdofcolorbox}
\def\dontdofcolorbox{\renewcommand\fcolorbox[4][]{##4}}

\else

\usepackage{fvextra}
\DefineVerbatimEnvironment{haskellcode}{Verbatim}{}
\DefineVerbatimEnvironment{Coqcode}{Verbatim}{}
\DefineVerbatimEnvironment{appendixcode}{Verbatim}{}

\fi

\BeforeBeginEnvironment{appendixcode}{\vspace{1em}}

\newenvironment{relation-rule}{
\noindent
\begin{minipage}{\textwidth}
  \vspace{0.5em}
  \begin{center}
  \begin{prooftree}
  }
  {
  \end{prooftree}
  \end{center}
  \vspace{0.5em}
\end{minipage}
}

\newcommand{\fom}{$F_{\omega}^{\mu}$}

\ifminted
\newcommand{\inlinecoq}[1]{\mintinline{Coq}{#1}}
\newcommand{\inlinehs}[1]{\mintinline{haskell}{#1}}
\else
\newcommand{\inlinecoq}[1]{\Verb{#1}}
\newcommand{\inlinehs}[1]{\Verb{#1}}
\fi

\newcommand{\altsep}[3]{~#1~& #2 & \text{#3}\\ }
\newcommand{\alt}[2]{\altsep{|}{#1}{#2}}

\newcommand{\lam}[3]{\lambda (#1 : #2).~ #3}
\newcommand{\Lam}[3]{\Uplambda (#1 : #2).~ #3}
\newcommand{\app}[2]{#1~ #2}
\newcommand{\tyapp}[2]{#1~ \{#2\}}

\newcommand{\keyword}[1]{\texttt{#1}~}

\newif\ifanon
\anonfalse

\begin{document}
\title{Translation Certification for Smart Contracts}
\ifanon
\author{Anonymous}
\else
\author{
  Jacco O.G. Krijnen
  \orcidlink{0000-0002-1840-472X}
  \inst{1}
  \and
  Manuel M. T. Chakravarty
  \inst{2}
  \and
  Gabriele Keller
  \orcidlink{0000-0003-1442-5387}
  \inst{1}
  \and
  Wouter Swierstra
  \orcidlink{0000-0002-0295-7944}
  \inst{1}
}
\fi

\ifanon
\authorrunning{Anonymous}
\else
\authorrunning{Krijnen et al.}
\fi

\ifanon
\else
\institute{
  Utrecht University\\
  \email{\{j.o.g.krijnen, w.s.swierstra, g.k.keller\}@uu.nl}
  \and
  IOHK\\
  \email{manuel.chakravarty@iohk.io}
  }
\fi

\maketitle

\begin{abstract}
Compiler correctness is an old problem, but with the emergence of \emph{smart contracts} on blockchains that problem presents itself in a new light. Smart contracts are self-contained pieces of software that control (valuable) assets in an adversarial environment; once committed to the blockchain, these smart contracts cannot be modified. Smart contracts are typically developed in a high-level contract language and compiled to low-level virtual machine code before being committed to the blockchain. For a smart contract user to trust a given piece of low-level code on the blockchain, they must convince themselves that (a) they are in possession of the matching source code and (b) that the compiler has correctly translated the source code to the given low-level code.

Classic approaches to compiler correctness tackle the second point. We argue that \emph{translation certification} also squarely addresses the first. We describe the proof architecture of a novel translation certification framework, implemented in Coq, for a functional smart contract language. We demonstrate that we can model the compilation pipeline as a sequence of translation relations that facilitate a modular verification methodology and are robust in the face of an evolving compiler implementation.

\end{abstract}

\section{Introduction}
\label{sec:introduction}

Compiler correctness is an old problem that has received renewed
interest in the context of \emph{smart contracts} --- that is,
compiled code on public blockchains, such as Ethereum or Cardano. This code
often controls a significant amount of financial assets, must operate under adversarial conditions, and can no
longer be updated once it has been committed to the blockchain. Bugs
in smart contracts are a significant problem in
practice~\cite{atzei-etal:survey}. Recent work has also established
that smart contract language compilers can exacerbate this
problem~\cite[Section~3]{park-etal:deposit-contract-verification} (in
this case, the Vyper compiler). More specifically, the authors report
(a) that they did find bugs in the Vyper compiler that compromised
smart contract security and (b) that they performed verification on
generated low-level code, because they were wary of compiler bugs.

Hence, to support reasoning about smart contract source code, we need
to get a handle on the correctness of smart contract compilers. On top
of that, we do also need a \emph{verifiable link} between the source
code and its compiled code to prevent \emph{code substitution
attacks,} where an adversary presents the user with source code that doesn't match the low-level code committed on-chain.

In this paper, we are reporting on our ongoing effort to develop a
certification engine for the open-source on-chain code compiler of the Plutus
smart contract
system\footnote{\url{https://developers.cardano.org/docs/smart-contracts/plutus/}} for
the Cardano blockchain.\footnote{\url{http://cardano.org} is, at the
time of writing, the 5th largest public blockchain by market
capitalisation.}
Specifically, we make the following contributions:
\begin{itemize}

\item

We describe a novel architecture for a translation certifier based on
\emph{translation relations,} which enables us to generate \emph{translation
certificates}---proof objects that relate the source code to the resulting
compiled code and establish the correctness of the translation
(Section~\ref{sec:architecture}).

\item

We provide formal definitions for the transformation passes that step-by-step
translate PIR (Plutus Intermediate Representation) to PLC (Plutus Core) and
briefly discuss the challenges associated with the certification of each of
these passes (Section~\ref{sec:plutus-passes}).

\item We present a summary of existing approaches to compiler
  correctness and discuss the importance of generating
  translation certificates in the domain of smart contracts
  (Section~\ref{sec:evaluation}).
\end{itemize}
We also evaluate how our approach to gradual certification copes with
changes to the compiler, which is being developed in an independent open source
project. Finally, we discuss related work in Section~\ref{sec:related-work} and
future work in Section~\ref{sec:future-work}.

\section{The Architecture of the Certifier}
\label{sec:architecture}

On-chain code in the Plutus smart contract system is written in a subset of Haskell called \emph{Plutus Tx} \cite{plutus:manual}. The Plutus Tx compiler is implemented as a plugin for
the widely-used, industrial-strength GHC Haskell compiler, combining
large parts of the GHC's compilation pipeline with custom translation
steps to generate Plutus Core. In this context, it seems infeasible to
apply full-scale compiler verification \`{a} la CompCert~\cite{leroy-etal:compcert}. We will therefore outline the design of a
certification engine that, using the Coq proof assistant \cite{barras:coq-manual,bertot-casteran:coqart}, generates a
proof object, a \emph{translation certificate}, asserting the validity
of a Plutus Core program with respect to a given Plutus Tx source
contract. In addition to asserting the correct translation \emph{of
this one program}, the translation certificate serves as a verifiable
link between source and generated code.

\begin{figure}[t]
\begin{center}
\includegraphics[width=.75\hsize]{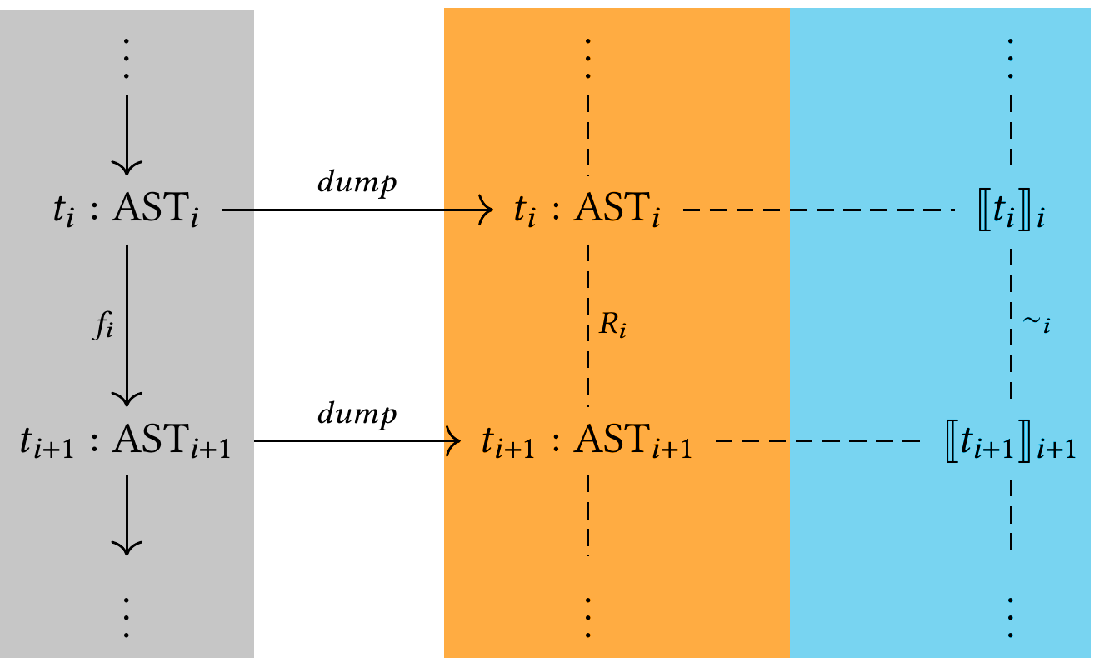}
\end{center}
\caption{Architecture for a single compiler pass. The grey area (left)
represents the compiler, orange (center) and blue (right) represent the
certification component in Coq.}
\label{fig:architecture}
\end{figure}

We model the compiler as a composition of pure functions that
transform one abstract syntax tree into another.
Figure~\ref{fig:architecture} illustrates the architecture for a
single transformation, where the grey area marks the compiler
implementation as a function $f_i : \text{AST}_i \to
\text{AST}_{i+1}$. We use a family of types $\text{AST}_i$ to
illustrate that the representation of the abstract syntax might change
after each transformation.

To support certification, the compiler outputs each intermediate tree
$t_i$, so that we can parse these in our Coq implementation of the
certifier. Within Coq, we define a high-level specification of each
pass. We call this specification a \emph{translation relation}: a binary relation on abstract syntax trees that specifies the intended behaviour of the compiler pass. The orange area in Figure~\ref{fig:architecture} displays
the translation relation $R_i$ of pass $i$, where the vertical dashed
line indicates that $R_i(t_i, t_{i+1})$ holds. To establish this, we
define a search procedure that, given two subsequent trees produced by the compiler, can
construct a derivation relating the two.

The translation relation is purely syntactic---it does not assert anything about the
correctness of the compiler---but rather \emph{specifies} the behaviour of a particular
compiler pass.
To verify that the compilation preserves language semantics requires an additional proof, the blue area
in Figure~\ref{fig:architecture}, that establishes that any two terms related by $R_i$
have the same semantics.

We have implemented this approach for a range of concrete passes of the Plutus Tx compiler. To illustrate our approach in this
section, we will use an untyped lambda calculus, extended with
non-recursive let-bindings.
\begin{align*}
  t ::= x ~|~  \lambda x.~t ~|~ \app{t}{t} ~|~ \texttt{let}~ x = t~ \texttt{in}~ t
\end{align*}
In the following section, we will extend this to a lambda calculus that is closer to the intermediate language used by the Plutus Tx compiler.

\subsection{Characterising a transformation}
\label{ss:translation-relation}

To assert the correctness of a single compiler stage $f_i$, we begin
by defining a translation relation $R_i$ on a pair of source and
target terms $t_i$ and $t_{i+1}$, respectively. This relation characterises
the admissible translations of that compiler stage. That is, for all $t_i,
t_{i+1}$, we have \(f_i(t_i) = t_{i+1}\text{ implies } R_i(t_i,t_{i+1})\).

As a concrete example, consider an inlining pass. We have characterised this as an
inductively defined relation in Figure~\ref{fig:inliner}. Here,
$\Upgamma \vdash s ~\triangleright ~t$ asserts that program $s$ can be
translated into $t$ given an environment $\Upgamma$ of let-bound
variables, paired with their definition.
According to Rule~[Inline-Var$_1$] the variable $x$ may be replaced by $t$
when the pair $(x, t')$ can be looked up in $\Upgamma$ and $t'$ can be translated
to $t$, accounting for repeated inlining. The remaining rules are congruence
rules, where Rule~[Inline-Let] also extends the environment $\Upgamma$.
We omitted details about handling variable capture to keep the
presentation simple: hence, we assume that variable names are
globally unique.

\begin{figure}[t]
\begin{center}

\begin{prooftree}
\hypo{\Upgamma(x) = t'}
\hypo{\Upgamma \vdash t' \triangleright t}
\infer2[[Inline-Var$_1$]]{\Upgamma \vdash x \triangleright t}
\end{prooftree}
\vspace{0.75em}

\begin{prooftree}
\infer0[[Inline-Var$_2$]]{\Upgamma \vdash x \triangleright x}
\end{prooftree}
\vspace{0.75em}

\begin{prooftree}
\hypo{\Upgamma \vdash t_1 \triangleright t_1'}
\hypo{(x, t_1), \Upgamma \vdash t_2 \triangleright t_2'}
\infer2[[Inline-Let]]
  {\Upgamma \vdash \textbf{let } x = t_1 \textbf{ in } t_2  \triangleright \textbf{let } x = t_1' \textbf{ in } t_2'}
\end{prooftree}
\vspace{0.75em}

\begin{prooftree}
\hypo{\Upgamma \vdash t_1 \triangleright t_1'}
\hypo{\Upgamma \vdash t_2 \triangleright t_2'}
\infer2[[Inline-App]]{\Upgamma \vdash t_1\ t_2 \triangleright t_1'\ t_2'}
\end{prooftree}
\vspace{0.75em}

\begin{prooftree}
\hypo{\Upgamma \vdash t_1 \triangleright t_1'}
\infer1[[Inline-Lam]]{\Upgamma \vdash \lambda x . t_1 \triangleright \lambda x . t_1'}
\end{prooftree}

\end{center}
\caption{Characterisation of an inliner}
\label{fig:inliner}
\end{figure}

Crucially, these rules do \emph{not} prescribe which variable
occurrences should be inlined, since the [Inline-Var$_1$] and [Inline-Var$_2$]
rules overlap. The choice in the implementation of the pass may rely
on a complex set of heuristics internal to the compiler. Instead, we
merely define a relation capturing the \emph{possible} ways in which
the compiler \emph{may} behave. This allows for a certification engine
that is robust with respect to changes in the compiler, such as the particular
heuristics used to decide when to replace a variable with its definition or not.

We can then encode the relation $\cdot \vdash \cdot \triangleright \cdot$ in
Coq as an inductive type \inlinecoq{Inline}, which is indexed by an environment
and two ASTs, as shown in Figure~\ref{fig:coq-inliner}. This type corresponds
closely to the rules of Figure \ref{fig:inliner}: we define exactly one
constructor per rule. However, there are some small differences. Since we
cannot omit details about variable capture anymore, we choose a de Bruijn
representation for variables and implement the environment $\Upgamma$ as a
cons-list. In the \inlinecoq{Inline_Let} constructor, we extend the list with
the bound term and furthermore shift free variables in the other bound terms.
For a let-bound variable $n$, its corresponding bound term can then be found at
the $n$'th position in the list using Coq's \inlinecoq{nth_error}
list-indexing function. For this indexing to work properly, the environment
also has to be extended at every lambda, as seen in \inlinecoq{Inline_Lam}. We
distinguish the two types of binding sites with the type \inlinecoq{binding}.

These inductive types implement the translation relation: its inhabitants are
proof derivations which will be a key ingredient of a compilation
certificate.

\begin{figure}[h!]
\begin{Coqcode}
Inductive binding :=
  | LetBound    : term -> binding
  | LambdaBound : binding.

Inductive Inline : list binding -> term -> term -> Type :=
  | Inline_Var_1 : forall {env n t},
      nth_error env n = Some (LetBound t) ->
      Inline env (Var n) t

  | Inline_Var_2 : forall {env n},
      Inline env (Var n) (Var n)

  | Inline_Let : forall {env s t s' t'},
      Inline env s s' ->
      Inline (LetBound s :: shiftEnv env) t t' ->
      Inline env (Let s t) (Let s' t')

  | Inline_Lam : forall {env s t},
      Inline (LambdaBound :: shiftEnv env) s t ->
      Inline env (Lam s) (Lam t)

  | Inline_App : forall {env s sx t tx},
      Inline env s t -> Inline env sx tx ->
      Inline env (App s sx) (App t tx)
  .
\end{Coqcode}
\caption{Characterisation of an inliner in Coq}
\label{fig:coq-inliner}
\end{figure}

\subsection{Proof search}

After defining a translation relation $R_i$ characterising one
compiler stage, we now define a decision procedure to construct a proof
that for two particular terms $t_i$ and $t_{i+1}$, produced by a run
of the compiler, the relation \(R_i(t_i, t_{i+1})\) holds. To find and implement
such a search procedure, we generally follow these steps:

\begin{enumerate}

\item

  We write proofs for specific compilations by hand using Coq's
  \emph{tactics}, a form of metaprogamming. For simple relations, like the inline example
  sketched above, a proof can often be found with a handful of tactics
  such as \inlinecoq{auto} or \inlinecoq{constructor}. This is
  particularly useful for debugging the design of our relations
  describing compiler passes. The drawback of this approach is,
  however, that it is difficult to reason when such proof search may
  fail. Furthermore, proofs written using such tactics quickly become
  slow for large terms.

\item

Once we are sufficiently confident that a relation accurately captures admissible compiler behaviour,
we write a decision procedure of the form \inlinecoq{forall (t1 t2 : term), option (R
t1 t2)}. These procedures can still produce large proof terms and may not
always successfully construct a proof, but they form a useful intermediate
step towards full-on proof by reflection.

\item

Finally, we write a boolean decision procedure in the style of
ssreflect~\cite{gonthier:ssreflect} of type \inlinecoq{term -> term -> bool},
together with a soundness proof stating that it will only return
\inlinecoq{true} when two terms are related through $R_i$. Verifying such
boolean functions for complex compilation passes is non-trivial; hence, we only
invest the effort once we have a reasonable degree of confidence that
the relation we have defined accurately describes a given compiler pass.

\end{enumerate}

\subsection{Semantics preservation}

Given the relational specification of each
individual compiler pass, we can now establish the correctness properties for
each pass. In the simplest case, this could be asserting the preservation of a
program's static semantics, i.e., a proof of type preservation. On the other
end of the spectrum, we can demonstrate that the translated term is
semantically equivalent to the original program. Proving such properties
for PIR and Plutus Core passes, however, requires advanced techniques such as step-indexed
logical relations~\cite{ahmed:LR}, which go beyond the scope of the
current paper.

In Figure~\ref{fig:architecture}, we denote $R_i$'s correctness properties in
the blue area by means of an abstract binary relation $\sim_i$ on the semantic
objects $\llbracket t_i \rrbracket_{i}$ of ASTs $t_i$. In the case of static
semantics, we can choose typing derivations as semantic objects, and (for most
passes) relate these by simply comparing types syntactically.

We can construct these proofs independently and gradually for each step in
the translation. In fact, even without any formal proof about the semantics,
inspection of the (relatively concise) definition of a translation relation may already provide some degree of confidence
that the translation step was performed correctly. After all, the translation relation
asserts the specification of this compiler pass' admissible behaviour.

\subsection{Certificate generation}

A complete translation certificate includes at least the entire set of ASTs \(t_1,
\ldots, t_n\) together with a proof term witnessing the translation
relations of type $R_1(t_1, t_2)\ \wedge\ \dots\ \wedge\ R_{n-1}(t_{n-1}, t_n)$. In addition,
any semantic preservation results on translation relations can be instantiated and 
included as a proof of $\llbracket t_i \rrbracket \sim_i \llbracket t_{i+1} \rrbracket$.

Together with the source and compiled program, one can now independently check the certificate
using a trusted proof checker, such as the Coq kernel~\cite{bertot-casteran:coqart}.
The definitions of the abstract syntax, translation relations and semantic preservation can be inspected
to confirm that the certificate proves the right theorem. One can then be confident that the
compiled program is a faithful translation of the source code.

\section{Translation Relations of the Plutus Tx Compiler}
\label{sec:plutus-passes}

The Plutus Tx compiler translates Plutus Tx (a subset of Haskell) to Plutus Core,
a variant of System~\fom~\cite{chapman-etal:system-f}. The
Plutus Core code is committed to the Cardano blockchain, constituting
the definitive reference to any deployed smart contract.

Plutus Core programs are pure, self-contained functions (i.e., they do not link
to other code) and are passed a representation of the transaction whose validation they contribute to. The programs are run by an interpreter during the transaction
validation phase of the blockchain.

The Plutus Tx compiler reuses parts of the GHC infrastructure and implements its custom passes by installing a core-to-core pass plugin~\cite{ghc:manual-plugins} in the GHC compiler pipeline. On a high level, the compiler comprises three steps:

\begin{enumerate}

\item The parsing, type-checking and desugaring phases of GHC are reused to
translate a surface-level Haskell program into a GHC Core program.

\item A large subset of GHC Core is directly translated into an intermediate
language named Plutus Intermediate Representation (PIR). These languages are similar and both based on
System F, with some extensions. Additionally, all referred
definitions are included as local definitions so that the program is self-contained.

\item\label{step:pir-plc} The PIR program is then transformed and compiled down
into Plutus Core.

\end{enumerate}
The certification effort reported here focuses on Step~\ref{step:pir-plc}, which
consists of several optimisation passes and translation steps. PIR is a
superset of the Plutus Core language: it adds several conveniences, such as
user-defined datatypes, strict and non-strict let-bindings that may be
(mutually) recursive. The compilation steps translate these constructs
into simpler language constructs.

\begin{figure}[t]
\begin{align*}
  t
    \altsep{::=}{x ~|~  \lam{x}{\tau}{t} ~|~ \app{t}{t} }
        {variable, lambda, function application}
    \alt{ \Lam{\alpha}{\kappa}{t} ~|~ \tyapp{t}{\tau}}
        {type abstraction, type application}
    \alt{\texttt{let}_r^s~ x = t~ \texttt{in}~ t}
        {term bindings}
    \alt{\texttt{data}~ T~ \overline{\alpha} = \overline{C_i~ \overline{\tau_i}}~ \keyword{with} x~ \keyword{in} t}
        {datatype binding}
    r
    \altsep{::=}{\keyword{rec} ~|~ \keyword{nonrec}}{recursion type of binding}
    s
    \altsep{::=}{\keyword{strict} ~|~ \keyword{nonstrict}}{strictness of binding}
    \tau
      \altsep{::=}{\dots}{types}
\end{align*}
\caption{Simplified PIR}
\label{def:pir-syntax}
\end{figure}
In Figure~\ref{def:pir-syntax} we present a simplified version of the
PIR syntax, where we omit some constructs for the sake of presentation. The full PIR language specification
has been formalised elsewhere~\cite{chapman-etal:system-f,jones-etal:unraveling}. In particular, we ignore
the fact that in PIR, let-bindings may contain a group of (mutually
recursive) bindings. Similarly, we do not include mutually-recursive
datatypes.  Furthermore, we omit the syntax of types, and the term-level
witnesses of iso-recursive types.
We occasionally omit type annotations, when they are not relevant.

We introduce the individual compiler passes that the Plutus Tx
compiler performs using the following Haskell program to illustrate
their behaviour:
\begin{haskellcode}
-- | Either a specific end date, or "never".
data EndDate = Fixed Integer | Never

pastEnd :: EndDate -> Integer -> Bool
pastEnd end current =
    let inlineMe = False
    in case end of
      Fixed n ->  (let floatMe = if current `greaterThanEqInteger` 0
        then n else 0 in floatMe) `lessThanEqInteger` current
      Never   -> inlineMe
\end{haskellcode}
This program is a basic implementation of a \emph{timelock}, a contract that
states that funds may be moved after a certain date, or not at all. It contains
a few contrived bindings (\inlinehs{inlineMe} and \inlinehs{floatMe}) that will be useful to illustrate some transformations. After the
program is desugared to GHC Core, it is converted to a term in PIR that corresponds to the following Simplified PIR term:

\begin{haskellcode}
data Bool = True | False with Bool_match in
  data Unit = Unit with Unit_match in
    let nonrec strict lessThanEqInteger = ... in
      data EndDate = Fixed Integer | Never with EndDate_match in
        \(end : EndDate).
        \(current : Integer).
          let nonrec nonstrict inlineMe = False in
          EndDate_match end
            (\unit n -> lessThanEqInteger
              (let nonrect nonstrict floatMe =
                Bool_match (greaterThanEqInteger current 0)
                  (\unit -> n) (\unit -> 0)
                  Unit
              in floatMe)
              current)
            (\unit -> inlineMe)
            Unit
\end{haskellcode}
Note that case distinction of a type \inlinehs{T} is encoded as the application
of a pattern match function \inlinehs{T_match}, which is introduced as part of
a data definition. Furthermore, branches of a
case distinction are delayed by abstracting over a unit value, since PIR is a strict language.

Next we will discuss the compiler passes, we have included each
intermediate form of the above program with some commentary in
\ifincludeappendix Appendix~\ref{app:appendix}. \else the appendix which can be found online\footnote{{\url{https://arxiv.org/abs/2201.04919}}}. \fi

\subsection{Variable Renaming}

\newcommand{\rename}{~\triangleright_\alpha}

In the renaming pass, the compiler transforms a program into an $\alpha$-equivalent
program, such that all variable names are globally unique, a property also known as the \emph{Barendregt-convention}. The implementation of some subsequent
compiler passes depend on it. We can express variable renaming as a
translation relation $\Updelta \vdash t \rename t'$, stating that under the
renaming environment $\Updelta$ (consisting of pairs of variables), $t$ is
renamed to $t'$. The environment $\Updelta$ records all variables that are
free in $t$, paired with their corresponding name in $t'$.

The case for lambda abstractions is defined as follows:

\vspace{0.5em}
\begin{relation-rule}
\hypo{(x, y), \Updelta \vdash t \rename t'}
\hypo{\{z ~|~ (z, y) \in \Updelta\} \cap FV(t) = \emptyset}
\infer2[[Rename-Abs]]{\Updelta \vdash \lambda x .
t \rename \lambda y . t'}
\end{relation-rule}

The [Rename-Abs] rule states that a lambda-bound variable $x$ may be renamed at its
binding-site to $y$, when $t$ and $t'$ are related under the extended environment. Of
course, $x$ may equal $y$, indicating that no renaming was performed.
Additionally, the new binder $y$ should not capture any other free variable $z$ in $t$ that was also
renamed to $y$. Very similar rules can be stated for other binding constructs
such as \inlinehs{let}.

Note that this relation does not establish global uniqueness of
variables: we consider that an implementation detail internal to the compiler. If this property
would be required or convenient in semantic preservation proofs, we will establish
it separately, allowing this renaming relation to be as general as possible.

The variable case simply follows from the environment $\Updelta$:

\begin{center}
\begin{prooftree}
\hypo{(x, y) \in \Updelta}
\infer1[[Rename-Var]]{\Updelta \vdash x \rename y}
\end{prooftree}
\end{center}

\subsection{Inlining}

The rules of the translation relation for inlining in PIR are similar to
those in Section~\ref{ss:translation-relation}. However, the Plutus Tx
compiler does more than just inlining let-bound definitions. It also performs
dead-code elimination (removing those let-bindings that have been inlined
exhaustively) and it renames variables to ensure the global uniqueness of bound
variables. This introduces a
problem for our certification approach, as we cannot observe and dump the
intermediate ASTs, since the transformations are fused into a single pass
in the compiler.

We solve this by modeling the individual transformations, composing them using \emph{relational composition}, $\exists t_2. R_1(t_1, t_2)
\land R_2(t_2, t_3)$. To construct a proof relating two terms, then amounts to
also finding the \emph{intermediate term}, $t_2$ witnessing the composite
transformation. To simplify the search of this intermediate AST, we
adjust the compiler to emit supporting information about the performed pass; in
this case, a list of the eliminated variables. If the compiler emits incorrect information, we may fail to construct a certificate, but we will never produce an incorrect certificate.

\subsection{Let-floating}
\label{sub:let-float}

\newcommand{\lfop}{\triangleright_{let}}
\newcommand{\lf}{~~\lfop~~}
\newcommand{\letin}[2]{\texttt{let}_r^\texttt{nonstrict}~#1~\texttt{in}~~#2}

During let-floating, let-bindings can be moved upwards in the program.
This may save unnecessarily repeated computation and makes the generated code
more readable.  The Plutus Tx compiler constructs a dependency graph to maintain a
correct ordering when multiple definitions are floated. For the translation
relation, we first consider the interaction of a \texttt{let} expression with
its parent node in the AST. For example, consider the case of a lambda with a
non-strict \texttt{let} directly under it:

\begin{relation-rule}
\hypo{x \notin FV(t_1)}
\hypo{x \neq y}
\hypo{t_1\lf t_1'}
\hypo{t_2 \lf t_2'}
\infer4[[Float-Let-Lam]]{\lambda x . \letin{y=t_1}{t_2}}
\infer[no rule]1{\lf}
\infer[no rule]1{\letin{y=t_1'}{\lambda x . t_2'} }
\end{relation-rule}
This rule states that a non-strict let-binding may float up past a lambda, if
the bound term does not reference the lambda-bound variable. Furthermore, we
require $x \neq y$, to avoid variable capture in $t_2$. This rule does not
apply to \texttt{strict} let-bindings, as floating them outside a lambda might
change termination behaviour of the program. Similar rules express when a
\texttt{let} may float upwards past the other language constructs. Most of these are
much simpler, only binding constructs pose additional constraints on scoping
and strictness. Since the compiler pass may float \texttt{let}s more than just one step
up, we define the translation relation as the transitive closure of $\lfop$.
Note that we do not need to maintain a dependency graph in the certifier, but only need to assert that transformations do not break dependencies.

\subsection{Dead-code elimination}

\newcommand{\dce}{~\triangleright_{dce}}

By means of a live variable analysis, the compiler determines which
let-bound definitions are unused. This is mainly useful for definitions that
are introduced by other compiler passes. Since PIR is a strict language,
however, the compiler can only eliminate those bindings for which it can
determine they have no side-effects. For example, a let-bound expression that
is unused but diverges cannot be removed, as that could change the termination
behaviour of the program.

The analysis in the compiler is not as straightforward as counting occurences. Even a let-bound variable that
does occur in the code, may be dead-code, if it is only used in other
dead bindings. This is also known as strongly live variable analysis~\cite{giegerich:deadcode}.
We define a translation relation $t \dce t'$ that captures dead code
elimination. The crucial rule is for let-bindings.
\begin{gather*}
 \dfrac
   {
   t_2 \dce t_2'
   \hspace{1em}
   x \notin FV(t_2')
   }
   {\textbf{let }_r^\text{nonstrict}~ x = t_1 \textbf{ in } t_2  \dce t_2'}
   \hspace{1em}
   \text{[DCE-Let-nonstrict]}
\end{gather*}
Note that the condition $x
\notin FV(t_2')$ mentions the \emph{resulting} body of the let $t_2'$. This is
justified since the rules of $\dce$ can remove bindings only, but cannot change
any other language constructs. This illustrates how succinct we can describe the specification
of a complex compiler pass.

In practice, the Plutus Tx compiler also eliminates some strict bindings that obviously do  not diverge, such as values.

\subsection{Encoding of non-strict bindings}

\newcommand{\thunk}{~\triangleright_{thunk}~}
\newcommand{\nonrec}{\texttt{nonrec}}
\newcommand{\rec}{\texttt{rec}}
\newcommand{\strict}{\texttt{strict}}
\newcommand{\nonstrict}{\texttt{nonstrict}}

The PIR language allows both for strict and non-strict let-bindings, but Plutus Core does not.
The \emph{thunking transformation} is used to obtain semantic equivalent definitions
which use a strict let-binding. We define the rules as a relation $\Upgamma \vdash t \thunk t'$, where
$\Upgamma$ records for every bound variable whether it was bound strictly or
non-strictly. The rule for a non-strict binding site is:
\begin{center}
\begin{prooftree}
  \hypo{ \Upgamma \vdash t_1 \thunk t_1'}
  \infer[no rule]1{ (x, \nonstrict), \Upgamma \vdash t_2 \thunk t_2' }
  \hypo{ y \notin FV(t_1) }
  \infer2[[Thunk-Let-nonstrict]]{
  \Upgamma \vdash
  }
  \infer[no rule]1{
  \textbf{let }_{\nonrec}^\nonstrict~ x = t_1 \textbf{ in } t_2
  }
  \infer[no rule]1{\thunk}
  \infer[no rule]1{
  \textbf{let }_{\nonrec}^\strict~ x = \lambda y.~ t_1' \textbf{ in } t_2'
  }
\end{prooftree}
\end{center}
This rule states that a right hand side is thunked by introducing a lambda abstraction
that expects a trivial unit value $y$ as its argument.

The rules for other variable binders extend $\Upgamma$.  The rule for a
recursive let-binding also extends the environment under which $t_1$
is transformed.  Finally, we also replace the occurrences of nonstrict variables, adding
an application to the unit value, thereby forcing evaluation.

\begin{center}
\begin{prooftree}
\hypo{(x, \nonstrict) \in \Upgamma}
\infer1[[Thunk-Var]]{\Upgamma \vdash x \thunk \app{x}{()}}
\end{prooftree}
\end{center}

\subsection{Encoding of recursive bindings}

\newcommand{\encrec}{~\triangleright_{\mu}~}
\newcommand{\fix}{fix~}

The Plutus Tx compiler translates (mutually) recursive let-bindings in
non-recursive ones using fixpoint combinators. Here we only consider the rule for individual recursive lets in simplified PIR:
\begin{center}

\begin{prooftree}
\hypo{t_1 \encrec t_1'}
\hypo{t_2 \encrec t_2'}
\hypo{y \notin FV(t_1)}
\infer3[[EncRec-Let]]{
\textbf{let }_{\rec}^s~ x = t_1 \textbf{ in } t_2
}
\infer[no rule]1{\encrec}
\infer[no rule]1{
    \textbf{let }_{\nonrec}^{\strict}~ \mathit{fix} = ... \textbf{ in }
    \textbf{let }_{\nonrec}^s~ x = \mathit{fix}~(\lambda x .~ t_1') \textbf{ in } t_2'
}
\end{prooftree}
\end{center}
\vspace{.5em}
This rule relates recursive bindings to non-recursive ones, and expects an
explicit definition of the fixpoint operator as well. Since PIR has no
primitive construct for term-level fix-points, the compiler generates a
definition $\mathit{fix}$. Note that $\mathit{fix}$ is defined in a non-recursive let, its
construction relies on recursive types~\cite{jones-etal:unraveling}.

The actual transformation for PIR is much more involved, since mutually
recursive binding groups require a more involved fixpoint combinator of which
the definition depends on the size of the group.

\subsection{Encoding of datatypes}

\newcommand{\encdata}{~\triangleright_{data}~}

Datatype definitions are encoded using lambda and type abstractions according
to the Scott encoding \cite{abadi:scott-numerals}. To show the idea of the
rather general $\encdata$ translation relation, we show a rule specialised to
the $\mathit{Maybe}$ datatype.

\begin{relation-rule}
\hypo{t \encdata t'}
\infer1[[Scott-Maybe]]{
\texttt{data}~\mathit{Maybe}~ \alpha = \mathit{Just}~ \alpha ~|~ \mathit{Nothing} ~\texttt{with}~ maybe ~\texttt{in}~ t
}
\infer[no rule]1{\encdata}
\infer[no rule]1{(\Uplambda \mathit{Maybe}. \lambda \mathit{Just}. \lambda \mathit{Nothing}. \lambda maybe.~ t')~ \tau_{\mathit{Maybe}}~ t_{\mathit{Just}}~ t_{\mathit{Nothing}}~ t_{maybe}}

\end{relation-rule}
%
%
The [Scott-Maybe] rule relates the datatype definition to a term that abstracts over the
type $\mathit{Maybe}$, its constructors $\mathit{Just}$ and $\mathit{Nothing}$ and the matching function
$maybe$, which are each lambda encoded. For the exact definitions of $\tau_{\mathit{Maybe}}$, $t_{\mathit{Just}}$, $t_{\mathit{Nothing}}$ and $t_{maybe}$
we refer to the general formalisation of PIR~\cite{jones-etal:unraveling}.

\subsection{Encoding of non-recursive bindings}

\newcommand{\desugar}{~\triangleright_{\beta}~}
A non-recursive let-binding is simply compiled into a $\beta$ redex:
\begin{gather*}
 \dfrac
   { t_1 \desugar t_1'
   \hspace{2em}
     t_2 \desugar t_2'
   }
   {
    \textbf{let }_{\nonrec}^{\strict}~ x = t_1 \textbf{ in } t_2
    \desugar
    \app{(\lambda x.~ t_2')}{t_1'}
   }
   \hspace{1em}
   \text{[Redex-Let]}
\end{gather*}
Note that at this point in the compiler pipeline, $\textbf{let
}_{\nonrec}^{\strict}$ is the only type of let-binding that can
still occur.

\section{Evaluation}
\label{sec:evaluation}

In this section, we evaluate our approach to proof engineering for an independently developed, constantly evolving compiler under the application constraints imposed by smart contracts.

\subsection{Compilers and correctness}

The standard approach to compiler correctness is \emph{full compiler verification}: a proof that asserts that the compiler is correct as it demonstrates that, for any valid source program, the translation produces a semantically equivalent target program. Examples of this approach include the CompCert~\cite{leroy-etal:compcert} and
CakeML~\cite{kumar-etal:cakeml} projects, showing that (with significant effort)
it is possible to verify a compiler end-to-end. To do so, the compiler is typically implemented in a language suitable for verification, such as the Coq proof
assistant or the HOL theorem prover.

In contrast, the technique that we propose for the Plutus Tx compiler is based on \emph{translation validation}~\cite{pnueli:translation-validation}. Instead of asserting an entire compiler correct, translation validation establishes the correctness of individual compiler runs.

A statement of full compiler correctness is, of course, the stronger of the two statements. Translation validation may fail to assert the correctness of some compiler runs; either because the compiler did not produce correct code or because the translation certifier is incomplete. In exchange for being the weaker property, translation validation is potentially (1) less costly to realise, (2) easier to retrofit to an existing compiler, and (3) more robust in the face of changes to the compiler.

The idea of \emph{proof-carrying code}~\cite{necula:pcc} is closely related to translation
validation, shifting the focus to compiled programs, rather than the
compiler itself. A program is distributed together with a proof of a property
such as memory or type safety. Such a proof excludes certain classes of bugs
and gives direct evidence to the users of such a program, who may independently
check the proof before running it. Our certification effort, while related, differs in that we keep proof and program separate and in that we are interested in full semantic correctness and not just certain properties like memory and type safety.

\subsection{Certificates and smart contracts}

Smart contracts often manage significant amounts of financial and other assets. Before a user engages with such a contract, which has been committed to the blockchain as compiled code, they may want to inspect the source code to assert that it behaves as they expect. In order to be able to rely on that inspection, they need to know without doubt that (1) they are looking at the correct source code and (2) that the source code has been  compiled correctly.

While a verified smart contract compiler addresses the second point, it doesn't help with the first. An infrastructure of \emph{reproducible builds}, on the other hand, solves only the first point. The latter is the approach taken by Etherscan\footnote{\url{https://etherscan.io/verifyContract}}: to verify that a deployed Ethereum smart contract was the result of a compiler run, one provides the source code and build information such as the compiler version and optimisation settings.

In contrast, a \emph{certifying compiler}~\cite{necula-lee:design-implementation} that generates an independently verifiable certificate of correct translation, squarely addresses both points. By verifying a smart contract's translation certificate, a smart contract user can convince themselves that they are in possession of the matching source code and that this was correctly compiled to the code committed to the blockchain.

\subsection{Engineering considerations}
\label{subsec:benefits}

\subsubsection{Gradual verification.}
The certifier architecture outlined in this paper allows for a gradual approach to
verification: during the development of the certification engine, each
individual step in the process increases our overall confidence in the
compiler's correctness, even if we have not yet completed the
end-to-end semantic verification of the compiler pipeline.

By defining only the translation relations, we have an independent formal
specification of the compiler's behaviour. This makes it easier to reason
informally and to spot potential mistakes or problems with the implementation.

Implementing the decision procedures for translation relations ties the
implementation to the specification: we can show on a per-compilation basis that a
pass is sound with respect to its specification as a translation relation. Furthermore, we can test and debug
translation relations by automatically constructing evidence for various input programs.

Finally, by proving semantics preservation of a translation relation, we gain full confidence in the corresponding pass for compiler runs that abide by that translation relation.

\subsubsection{Agility.}
The Plutus Tx compiler is developed independently of our certification effort. Moreover, it depends on large parts of a large code base --- namely, that of the Glasgow Haskell Compiler (GHC). In addition, both GHC and the Plutus Tx-specific parts evolve on a constant basis; for example, to improve code optimisation or to fix bugs.

In that context, full verification appears an insurmountable task and a proof on the basis of the compiler source code would constantly have to adapt to the evolving compiler source. Hence, the architecture of our certification engine is based on a \emph{grey box approach}, where the certifier matches the general outline (such as the phases of the compiler pipeline), but not all of the implementation details of the compiler. For example, our translation relation for the inliner admits any valid inlining.
Improvements of the compiler heuristics to produce more efficient programs by
being selective about what precisely to inline don't affect the inliner's
translation relation, and hence, don't affect the certifier.

\subsubsection{Trusted Computing Base (TCB).}

The fact that the Plutus Tx compiler is not implemented in a proof assistant, but in Haskell complicates direct compiler verification. It might be possible to
use a tool like hs-to-coq~\cite{spector:hs-to-coq},
which translates a subset of Haskell into Coq's Gallina and has been used for
proving various properties about Haskell code~\cite{breitner:hs-to-coq}.
However, given that those tools often only cover language subsets, it is not clear that they are applicable. More importantly,
such an approach would increase the size of the trusted computing base (TCB), as the translation from Haskell into Coq's Gallina is not verified. Similarly, extraction-based approaches suffer from the same problem if the extraction itself is not verified, although there are projects like CertiCoq~\cite{anand-etal:certicoq} that try to address that issue.

In any case, our architecture has a small TCB. We directly relate the source
and target programs, taking the compiler implementation out of the equation. Trusting a translation certificate comes down to trusting the Coq
kernel that checks the proof, the theorem with its supporting definitions and
soundness of the Plutus Core interpreter with respect to the formalised semantics.
Of course, these components are part of the TCB of a verified compiler too.
This aspect also motivated our choice of Coq over other languages such as Agda,
due to its relatively small and mature kernel.

\section{Related Work}
\label{sec:related-work}

Ethereum was the first blockchain to popularise use of smart contracts, written
in the Solidity programming language. Solidity is an imperative programming
language that is compiled to EVM bytecode, which runs on a stack machine
operating on persistent mutable state. The DAO
vulnerability~\cite{buterin:thedao} has underlined the importance of formal
verification of smart contracts. Notably, a verification framework has
been presented ~\cite{bhargavan:solidity} for reasoning about embedded Solidity
programs in F*. The work includes a decompiler to convert EVM bytecode, generated by
a compiler, into Solidity programs in F*. The authors propose that correctness
of compilation can be shown by proving equivalence of the embedded source and
(decompiled) target program using relational
reasoning~\cite{barthe:relational-verif}. However, this would involve a manual
proof effort on a per-program basis, and relies on the F* semantics since the
embeddings are shallow. Furthermore, components such as the decompiler are not
formally verified, adding to the size of the TCB.

The translation validation technique has been used for the verification of a
particular critical Ethereum smart
contract~\cite{park-etal:deposit-contract-verification} using the K framework.
The work demonstrates how translation validation can succesfully be applied to
construct proofs about the low-level EVM bytecode by mostly reasoning on the
(much more understandable) source code. The actual refinement proof is still
constructed manually, however.

The Tezos blockchain also uses a stack-like language, called Michelson. The
Mi-Cho-Coq framework~\cite{bernardo:michelson} formalises the language and supports reasoning with
a weakest precondition logic. There is ongoing work for developing a
certified compiler in Coq for the Albert intermediate language, intended as a target language
for certified compilers of higher-level languages. This differs from
our approach as it requires the compiler to be implemented in the proof
assistant.

ConCert is a smart contract verification framework in Coq~\cite{annenkov:concert}. It enables
formal reasoning about the source code of a smart contracts, defined in a
different (functional) language. The programs are translated and shallowly
embedded in Coq's Gallina. Interestingly, the translation is proven sound, in
contrast with approaches such as hs-to-coq~\cite{spector:hs-to-coq}, since it is implemented
using Coq's metaprogramming and reasoning facility MetaCoq~\cite{sozeau:metacoq}.

The Cogent certifying compiler~\cite{liam:cogent} has shown that it is possible
to use translation validation for lowering the cost of functional verification
of low-level code: a program can be written and reasoned about in a high-level
functional language, which is compiled down to C. The generated certificate
then proves a refinement relation, capable of transporting the verification
results to the corresponding C code. The situation is different from
ours: the Cogent compiler goes through a range of languages with
different semantic models and uses the forward-simulation technique as a consequence.
In contrast, we are working with variations of lambda calculi that have similar
semantics, allowing us to use logical relations and translation relations.

In their Coq framework~\cite{li:rewrite-rules}, Li and Appel use a similar
technique for specifying compiler passes as inductive relations in Coq. Their
tool reduces the effort of implementing program transformations and corresponding
correctness proofs. The tool is able to generate large parts of an
implementation together with a partial soundess proof with respect to those
relations. The approach is used to implement parts of the CertiCoq backend.

\section{Conclusions and further work}
\label{sec:future-work}

The Plutus Tx compiler translates a Haskell subset into Plutus Core. The compiler
consists of three main parts: the first one reuses various stages of GHC to
compile the Haskell subset to GHC Core --- GHC's principal intermediate
language. The second part translates GHC Core to PIR and the final part compiles PIR to Plutus Core. As Plutus Core is
strict and doesn't directly support datatypes, these parts are quite complex.
Moreover, they consist of a significant number of successive transformation
steps.

In this paper, we focused on the certification effort covering the third part of that pipeline; specifically, the translation steps from PIR to Plutus Core. We developed
translation relations for all passes described in Section~\ref{sec:plutus-passes}, such that we can, for example, produce a proof relating the previously described timelock example in PIR to
its final form in Plutus Core. For some of these passes, such as inlining, we
have implemented a verified decision procedure, but most of the evidence is
generated semi-automatically by using Coq tactics. We have not yet covered all
transformations in their full generality; for example, we do not cover (mutually) recursive datatypes yet. We have also started the semantic verification of key
passes of the translation\cite{dral:thesis} and are investigating different ways to improve the efficiency of proof search for larger programs.

Our next steps comprise the following: (1) filling in the remaining gaps in translation relations (such as covering mutually recursive datatypes); (2) complete all decision procedures; (3) drive the semantic verification forward; and (4) develop techniques to further automate our approach and improve the efficiency of the certifier.

The first three steps pose a significant amount of work, but we do not expect major new conceptual questions or obstacles. This is different for Step (4), where we anticipate the need for further research work. This includes more compositional definitions of the translation relations, such that we can generate at least part of the decision procedures (semi-)automatically. Moreover, we already perceive efficiency to be a bottleneck and we plan to work on optimising the proof search. Finally, we plan to apply our approach to the first part of the Plutus Tx compiler (Haskell subset to GHC Core).

\bibliographystyle{splncs04}
\bibliography{references}

\ifincludeappendix
\newpage
\appendix

\newenvironment{changemargin}[2]{%
\begin{list}{}{%
\setlength{\topsep}{0pt}%
\setlength{\leftmargin}{#1}%
\setlength{\rightmargin}{#2}%
\setlength{\listparindent}{\parindent}%
\setlength{\itemindent}{\parindent}%
\setlength{\parsep}{\parskip}%
}%
\item[]}{\end{list}}

\begin{changemargin}{-1.5cm}{-1cm}
\section{Compiler dumps for the timelock program}
\label{app:appendix}

In this appendix we show step-by-step how the timelock example in section
\ref{sec:plutus-passes} is transformed by the passes in the Plutus Tx compiler. These programs
were obtained by running the Plutus Tx compiler on the Haskell source code
program, after modifying the pretty-printer to output a bit more compact
presentation. We ocassionally omit some sub-terms to improve readability
(indicated as \texttt{...}).

\subsection{Original PIR Term}
\label{app:pir-start}

The Plutus Tx compiler converts the GHC Core program into the following PIR
program. Note that variables in PIR are represented as pairs of names and
unique integers. The name is only maintained for readability, whereas the
integers are used for actual program transformations. We pretty-print the
integer in subscript after the name.

The conversion includes definitions for all the built-in types and functions
that may be used in PIR program, since the program has to be self-contained.
Starting from line 34 we can recognise the timelock example. Note that
Haskell's lazy \inlinehs{case} expression has been translated to a call to
\inlinehs{EndDate_match}, where the case branches have been ``thunked'' by abstracting
over a unit value. This thunking prevents the (strict) function application of
\inlinehs{EndDate_match} from evaluating all the branches.

\begin{appendixcode}
let nonrec type ByteString@$_{0}$@ = ... in
let nonrec data Bool@$_{11}$@  = True@$_{13}$@ : ... | False@$_{14}$@ : ... with Bool_match@$_{12}$@ in
let nonrec strict verifySignature@$_{57}$@ = ... in
let nonrec type String@$_{2}$@ = ... in
let nonrec data Unit@$_{60}$@  = Unit@$_{62}$@ : ... with Unit_match@$_{61}$@ in
let nonrec strict trace@$_{70}$@ = ... in
let nonrec type Integer@$_{1}$@ = ... in
let nonrec strict takeByteString@$_{5}$@ = ... in
let nonrec strict subtractInteger@$_{27}$@ = ... in
let nonrec strict sha3_@$_{8}$@ = ... in
let nonrec strict sha2_@$_{7}$@ = ... in
let nonrec strict remainderInteger@$_{32}$@ = ... in
let nonrec strict quotientInteger@$_{31}$@ = ... in
let nonrec strict multiplyInteger@$_{28}$@ = ... in
let nonrec strict modInteger@$_{30}$@ = ... in
let nonrec strict lessThanInteger@$_{44}$@ = ... in
let nonrec strict lessThanEqInteger@$_{48}$@ = ... in
let nonrec strict lessThanByteString@$_{20}$@ = ... in
let nonrec strict greaterThanInteger@$_{36}$@ = ... in
let nonrec strict greaterThanEqInteger@$_{40}$@ = ... in
let nonrec strict greaterThanByteString@$_{24}$@ = ... in
let nonrec strict error@$_{64}$@ = ... in
let nonrec strict equalsInteger@$_{52}$@ = ... in
let nonrec strict equalsByteString@$_{16}$@ = ... in
let nonrec strict emptyString@$_{66}$@ = ... in
let nonrec strict emptyByteString@$_{25}$@ = ... in
let nonrec strict dropByteString@$_{6}$@ = ... in
let nonrec strict divideInteger@$_{29}$@ = ... in
let nonrec strict concatenate@$_{4}$@ = ... in
let nonrec type Char@$_{3}$@ = ... in
let nonrec strict charToString@$_{67}$@ = ... in
let nonrec strict appendString@$_{65}$@ = ... in
let nonrec strict addInteger@$_{26}$@ = ... in
let nonrec data EndDate@$_{71}$@  = Fixed@$_{73}$@ : ... | Never@$_{74}$@ : ... with EndDate_match@$_{72}$@ in
λds@$_{75}$@ : EndDate@$_{71}$@ .
λds@$_{76}$@ : Integer .
let nonrec nonstrict inlineMe@$_{77}$@ = False in
let nonrec nonstrict wild@$_{78}$@ = ... in
(((EndDate_match@$_{72}$@ ds@$_{75}$@ { Unit@$_{60}$@ -> Bool@$_{11}$@ }) (λn@$_{79}$@ : Integer .
λthunk@$_{84}$@ : Unit@$_{60}$@ .
(lessThanEqInteger@$_{48}$@ (let nonrec nonstrict floatMe@$_{83}$@ = ... in
floatMe@$_{83}$@)) ds@$_{76}$@)) (λthunk@$_{85}$@ : Unit@$_{60}$@ .
inlineMe@$_{77}$@)) Unit@$_{62}$@
\end{appendixcode}

\subsection{Renaming}
\label{app:renamer}

The first pass does a global renaming to ensure each variable is in fact
globally unique. Note that the integers of all bound variables are indeed
renamed compared to the previous version of the program.

\begin{appendixcode}
let nonrec type ByteString@$_{86}$@ = ... in
let nonrec data Bool@$_{87}$@  = True@$_{88}$@ : ... | False@$_{89}$@ : ... with Bool_match@$_{90}$@ in
let nonrec strict verifySignature@$_{91}$@ = ... in
let nonrec type String@$_{96}$@ = ... in
let nonrec data Unit@$_{97}$@  = Unit@$_{98}$@ : ... with Unit_match@$_{99}$@ in
let nonrec strict trace@$_{100}$@ = ... in
let nonrec type Integer@$_{103}$@ = ... in
let nonrec strict takeByteString@$_{104}$@ = ... in
let nonrec strict subtractInteger@$_{105}$@ = ... in
let nonrec strict sha3_@$_{106}$@ = ... in
let nonrec strict sha2_@$_{107}$@ = ... in
let nonrec strict remainderInteger@$_{108}$@ = ... in
let nonrec strict quotientInteger@$_{109}$@ = ... in
let nonrec strict multiplyInteger@$_{110}$@ = ... in
let nonrec strict modInteger@$_{111}$@ = ... in
let nonrec strict lessThanInteger@$_{112}$@ = ... in
let nonrec strict lessThanEqInteger@$_{116}$@ = ... in
let nonrec strict lessThanByteString@$_{120}$@ = ... in
let nonrec strict greaterThanInteger@$_{124}$@ = ... in
let nonrec strict greaterThanEqInteger@$_{128}$@ = ... in
let nonrec strict greaterThanByteString@$_{132}$@ = ... in
let nonrec strict error@$_{136}$@ = ... in
let nonrec strict equalsInteger@$_{140}$@ = ... in
let nonrec strict equalsByteString@$_{144}$@ = ... in
let nonrec strict emptyString@$_{148}$@ = ... in
let nonrec strict emptyByteString@$_{149}$@ = ... in
let nonrec strict dropByteString@$_{150}$@ = ... in
let nonrec strict divideInteger@$_{151}$@ = ... in
let nonrec strict concatenate@$_{152}$@ = ... in
let nonrec type Char@$_{153}$@ = ... in
let nonrec strict charToString@$_{154}$@ = ... in
let nonrec strict appendString@$_{155}$@ = ... in
let nonrec strict addInteger@$_{156}$@ = ... in
let nonrec data EndDate@$_{157}$@  = Fixed@$_{158}$@ : ... | Never@$_{159}$@ : ... with EndDate_match@$_{160}$@ in
λds@$_{161}$@ : EndDate@$_{157}$@ .
λds@$_{162}$@ : Integer .
let nonrec nonstrict inlineMe@$_{163}$@ = False in
let nonrec nonstrict wild@$_{164}$@ = ... in
(((EndDate_match@$_{160}$@ ds@$_{161}$@ { Unit@$_{97}$@ -> Bool@$_{87}$@ }) (λn@$_{165}$@ : Integer .
λthunk@$_{166}$@ : Unit@$_{97}$@ .
(lessThanEqInteger@$_{116}$@ (let nonrec nonstrict floatMe@$_{167}$@ = ... in
floatMe@$_{167}$@)) ds@$_{162}$@)) (λthunk@$_{171}$@ : Unit@$_{97}$@ .
inlineMe@$_{163}$@)) Unit@$_{98}$@
\end{appendixcode}

\subsection{Dead Code Elimination}
\label{app:dce}

In this pass, the compiler cleans up the unused definitions that were present after the GHC core translation.

\begin{appendixcode}
let nonrec data Bool@$_{1}$@  = True@$_{2}$@ : ... | False@$_{3}$@ : ... with Bool_match@$_{4}$@ in
let nonrec data Unit@$_{11}$@  = Unit@$_{12}$@ : ... with Unit_match@$_{13}$@ in
let nonrec strict lessThanEqInteger@$_{30}$@ = ... in
let nonrec strict greaterThanEqInteger@$_{42}$@ = ... in
let nonrec data EndDate@$_{71}$@  = Fixed@$_{72}$@ : ... | Never@$_{73}$@ : ... with EndDate_match@$_{74}$@ in
λds@$_{75}$@ : EndDate@$_{71}$@ .
λds@$_{76}$@ : Integer .
let nonrec nonstrict inlineMe@$_{77}$@ = False in
(((EndDate_match@$_{74}$@ ds@$_{75}$@ { Unit@$_{11}$@ -> Bool@$_{1}$@ }) (λn@$_{79}$@ : Integer .
λthunk@$_{80}$@ : Unit@$_{11}$@ .
(lessThanEqInteger@$_{30}$@ (let nonrec nonstrict floatMe@$_{81}$@ = ... in
floatMe@$_{81}$@)) ds@$_{76}$@)) (λthunk@$_{85}$@ : Unit@$_{11}$@ .
inlineMe@$_{77}$@)) Unit@$_{12}$@
\end{appendixcode}

\subsection{Inlining}
\label{app:inliner}

The compiler performs an inlining pass and decides to inline the let-bound
definition \inlinehs{inlineMe} on line 8 in Section \ref{app:dce}. This results in the following program,
where the let-binding has been eliminated and the inlined definition (\inlinehs{False}) can be seen
on line 12.

\begin{appendixcode}
let nonrec data Bool@$_{1}$@  = True@$_{2}$@ : ... | False@$_{3}$@ : ... with Bool_match@$_{4}$@ in
let nonrec data Unit@$_{11}$@  = Unit@$_{12}$@ : ... with Unit_match@$_{13}$@ in
let nonrec strict lessThanEqInteger@$_{30}$@ = ... in
let nonrec strict greaterThanEqInteger@$_{42}$@ = ... in
let nonrec data EndDate@$_{71}$@  = Fixed@$_{72}$@ : ... | Never@$_{73}$@ : ... with EndDate_match@$_{74}$@ in
λds@$_{75}$@ : EndDate@$_{71}$@ .
λds@$_{76}$@ : Integer .
(((EndDate_match@$_{74}$@ ds@$_{75}$@ { Unit@$_{11}$@ -> Bool@$_{1}$@ }) (λn@$_{79}$@ : Integer .
λthunk@$_{80}$@ : Unit@$_{11}$@ .
(lessThanEqInteger@$_{30}$@ (let nonrec nonstrict floatMe@$_{81}$@ = ... in
floatMe@$_{81}$@)) ds@$_{76}$@)) (λthunk@$_{85}$@ : Unit@$_{11}$@ .
False@$_{3}$@)) Unit@$_{12}$@
\end{appendixcode}

\subsection{Thunking recursive definitions}
\label{app:thunk-rec}

The next pass thunks recursive term bindings (similar to the encoding of non-strict let bindings in
Section \ref{app:let-non-strict}), to make sure that they
are of a function type and work well with the fixpoint combinator
that is introduced in a later pass (Section \ref{app:let-rec}). Since this
program does not include any recursive term bindings, the result is
unchanged.

\subsection{Let-floating}
\label{app:let-float}

Next, the Plutus Tx compiler decides to float a let-bound definition. In this run,
the \inlinehs{floatMe} definition is moved outside of the first argument of
\inlinehs{lessThanEqInteger}, as can be seen on line 10. Additionally, this
pass performs \emph{merging} of adjacent let definitions into a single let with
a group of bindings, printed on line 1. We did not mention this transformation
in Section \ref{sub:let-float}, since simplified PIR has no binding groups.
The order of these definitions has also changed, but this is fine as long as no
dependencies are broken. We use a translation relation that is reminiscent of the
one in Section{sub:let-float}, but for bindings only.

\begin{appendixcode}
let nonrec data Bool@$_{1}$@  = True@$_{2}$@ : ... | False@$_{3}$@ : ... with Bool_match@$_{4}$@;
    strict greaterThanEqInteger@$_{42}$@ = ...;
    data Unit@$_{11}$@  = Unit@$_{12}$@ : ... with Unit_match@$_{13}$@;
    data EndDate@$_{71}$@  = Fixed@$_{72}$@ : ... | Never@$_{73}$@ : ... with EndDate_match@$_{74}$@;
    strict lessThanEqInteger@$_{30}$@ = ... in
λds@$_{75}$@ : EndDate@$_{71}$@ .
λds@$_{76}$@ : Integer .
(((EndDate_match@$_{74}$@ ds@$_{75}$@ { Unit@$_{11}$@ -> Bool@$_{1}$@ }) (λn@$_{79}$@ : Integer .
λthunk@$_{80}$@ : Unit@$_{11}$@ .
let nonrec nonstrict floatMe@$_{81}$@ = ... in
(lessThanEqInteger@$_{30}$@ floatMe@$_{81}$@) ds@$_{76}$@)) (λthunk@$_{85}$@ : Unit@$_{11}$@ .
False@$_{3}$@)) Unit@$_{12}$@
\end{appendixcode}

\subsection{Encoding of non-strict let bindings}
\label{app:let-non-strict}

The non-strict binding on line 10 is transformed in a strict binding by
thunking. From the type we can see that the Plutus Tx compiler actually abstracts
over the Scott-encoded version of a unit value. The occurrence is
applied to a unit value on line 11.

\begin{appendixcode}
let nonrec data Bool@$_{1}$@  = True@$_{2}$@ : ... | False@$_{3}$@ : ... with Bool_match@$_{4}$@;
    strict greaterThanEqInteger@$_{42}$@ = ...;
    data Unit@$_{11}$@  = Unit@$_{12}$@ : ... with Unit_match@$_{13}$@;
    data EndDate@$_{71}$@  = Fixed@$_{72}$@ : ... | Never@$_{73}$@ : ... with EndDate_match@$_{74}$@;
    strict lessThanEqInteger@$_{30}$@ = ... in
λds@$_{75}$@ : EndDate@$_{71}$@ .
λds@$_{76}$@ : Integer .
(((EndDate_match@$_{74}$@ ds@$_{75}$@ { Unit@$_{11}$@ -> Bool@$_{1}$@ }) (λn@$_{79}$@ : Integer .
λthunk@$_{80}$@ : Unit@$_{11}$@ .
let nonrec strict floatMe@$_{81}$@ = λarg@$_{207}$@ : ∀a@$_{0}$@ : *.a@$_{0}$@ -> a@$_{0}$@ . ... in
(lessThanEqInteger@$_{30}$@ (floatMe@$_{81}$@ (Λa@$_{0}$@ : *.
λx@$_{1}$@ : a@$_{0}$@ .
x@$_{1}$@))) ds@$_{76}$@)) (λthunk@$_{85}$@ : Unit@$_{11}$@ .
False@$_{3}$@)) Unit@$_{12}$@
\end{appendixcode}

\subsection{Encoding of datatypes}
\label{app:scott}

Next, the three datatype definitions are Scott encoded. For example, the
\inlinehs{Bool} datatype with its constructors and matching function are bound on
line 1 to 4, and the corresponding definitions are found as arguments on line
40 for \inlinehs{Bool}, line 41-44 for \inlinehs{True}, line 45-48 for
\inlinehs{False} and line 49-50 for \inlinehs{Bool_match}.

\begin{appendixcode}
(((ΛBool@$_{1}$@ : *.
λTrue@$_{2}$@ : Bool@$_{1}$@ .
λFalse@$_{3}$@ : Bool@$_{1}$@ .
λBool_match@$_{4}$@ : Bool@$_{1}$@ -> (∀a@$_{221}$@ : *.a@$_{221}$@ -> (a@$_{221}$@ -> a@$_{221}$@)) .
let nonrec strict greaterThanEqInteger@$_{42}$@ = ... in
((ΛUnit@$_{11}$@ : *.
λUnit@$_{12}$@ : Unit@$_{11}$@ .
λUnit_match@$_{13}$@ : Unit@$_{11}$@ -> (∀a@$_{217}$@ : *.a@$_{217}$@ -> a@$_{217}$@) .
(((ΛEndDate@$_{71}$@ : *.
λFixed@$_{72}$@ : Integer -> EndDate@$_{71}$@ .
λNever@$_{73}$@ : EndDate@$_{71}$@ .
λEndDate_match@$_{74}$@ : EndDate@$_{71}$@ -> (∀a@$_{208}$@ : *.(Integer -> a@$_{208}$@) -> (a@$_{208}$@ -> a@$_{208}$@)) .
let nonrec strict lessThanEqInteger@$_{30}$@ = ... in
λds@$_{75}$@ : EndDate@$_{71}$@ .
λds@$_{76}$@ : Integer .
(((EndDate_match@$_{74}$@ ds@$_{75}$@ { Unit@$_{11}$@ -> Bool@$_{1}$@ }) (λn@$_{79}$@ : Integer .
λthunk@$_{80}$@ : Unit@$_{11}$@ .
let nonrec strict floatMe@$_{81}$@ = ... in
(lessThanEqInteger@$_{30}$@ (floatMe@$_{81}$@ (Λa@$_{0}$@ : *.
λx@$_{1}$@ : a@$_{0}$@ .
x@$_{1}$@))) ds@$_{76}$@)) (λthunk@$_{85}$@ : Unit@$_{11}$@ .
False@$_{3}$@)) Unit@$_{12}$@ { ∀a@$_{208}$@ : *.(Integer -> a@$_{208}$@) -> (a@$_{208}$@ -> a@$_{208}$@) })
(λarg_0@$_{212}$@ : Integer .
   Λa@$_{209}$@ : *.
   λcase_Fixed@$_{210}$@ : Integer -> a@$_{209}$@ .
   λcase_Never@$_{211}$@ : a@$_{209}$@ .
    case_Fixed@$_{210}$@ arg_0@$_{212}$@))
(Λa@$_{213}$@ : *.
  λcase_Fixed@$_{214}$@ : Integer -> a@$_{213}$@ .
  λcase_Never@$_{215}$@ : a@$_{213}$@ .
    case_Never@$_{215}$@))
(λx@$_{216}$@ : ∀a@$_{208}$@ : *.(Integer -> a@$_{208}$@) -> (a@$_{208}$@ -> a@$_{208}$@) .
  x@$_{216}$@)
{ ∀a@$_{217}$@ : *.a@$_{217}$@ -> a@$_{217}$@ })
(Λa@$_{218}$@ : *.
  λcase_Unit@$_{219}$@ : a@$_{218}$@ .
    case_Unit@$_{219}$@))
(λx@$_{220}$@ : ∀a@$_{217}$@ : *.a@$_{217}$@ -> a@$_{217}$@ .
  x@$_{220}$@)
{ ∀a@$_{221}$@ : *.a@$_{221}$@ -> (a@$_{221}$@ -> a@$_{221}$@) })
(Λa@$_{222}$@ : *.
  λcase_True@$_{223}$@ : a@$_{222}$@ .
  λcase_False@$_{224}$@ : a@$_{222}$@ .
  case_True@$_{223}$@))
(Λa@$_{225}$@ : *.
  λcase_True@$_{226}$@ : a@$_{225}$@ .
  λcase_False@$_{227}$@ : a@$_{225}$@ .
    case_False@$_{227}$@))
(λx@$_{228}$@ : ∀a@$_{221}$@ : *.a@$_{221}$@ -> (a@$_{221}$@ -> a@$_{221}$@) .
  x@$_{228}$@)
\end{appendixcode}

\subsection{Recursive let bindings, inlining and dead-code elimination}
\label{app:let-rec}

The next three passes encode recursive term-bindings, and perform another round
of inlining and dead code elimination.  In this example program however, they
have no effect and the program does not change.

\subsection{Non-recursive let bindings}
\label{app:let-nonrec}

The final pass encodes non-recursive let bindings as a beta-redex. The
\inlinehs{floatMe} binding in Section \ref{app:scott} line 18 can be recognised
below on line 18, where it is now lambda-bound, and line 43 where the definition
is provided as an argument.

\begin{appendixcode}
(((ΛBool@$_{0}$@ : *.
λTrue@$_{1}$@ : Bool@$_{0}$@ .
λFalse@$_{2}$@ : Bool@$_{0}$@ .
λBool_match@$_{3}$@ : Bool@$_{0}$@ -> (∀a@$_{4}$@ : *.a@$_{4}$@ -> (a@$_{4}$@ -> a@$_{4}$@)) .
(λgreaterThanEqInteger@$_{5}$@ : Integer -> (Integer -> Bool@$_{0}$@) .
((ΛUnit@$_{9}$@ : *.
λUnit@$_{10}$@ : Unit@$_{9}$@ .
λUnit_match@$_{11}$@ : Unit@$_{9}$@ -> (∀a@$_{12}$@ : *.a@$_{12}$@ -> a@$_{12}$@) .
(((ΛEndDate@$_{13}$@ : *.
λFixed@$_{14}$@ : Integer -> EndDate@$_{13}$@ .
λNever@$_{15}$@ : EndDate@$_{13}$@ .
λEndDate_match@$_{16}$@ : EndDate@$_{13}$@ -> (∀a@$_{17}$@ : *.(Integer -> a@$_{17}$@) -> (a@$_{17}$@ -> a@$_{17}$@)) .
(λlessThanEqInteger@$_{18}$@ : Integer -> (Integer -> Bool@$_{0}$@) .
λds@$_{22}$@ : EndDate@$_{13}$@ .
λds@$_{23}$@ : Integer .
(((EndDate_match@$_{16}$@ ds@$_{22}$@ { Unit@$_{9}$@ -> Bool@$_{0}$@ }) (λn@$_{24}$@ : Integer .
λthunk@$_{25}$@ : Unit@$_{9}$@ .
(λfloatMe@$_{26}$@ : (∀a@$_{27}$@ : *.a@$_{27}$@ -> a@$_{27}$@) -> Integer .
(lessThanEqInteger@$_{18}$@ (floatMe@$_{26}$@ (Λa@$_{32}$@ : *.
λx@$_{33}$@ : a@$_{32}$@ .
x@$_{33}$@))) ds@$_{23}$@) (λarg@$_{28}$@ : ∀a@$_{29}$@ : *.a@$_{29}$@ -> a@$_{29}$@ .
(((Bool_match@$_{3}$@ ((greaterThanEqInteger@$_{5}$@ ds@$_{23}$@) 0) { Unit@$_{9}$@ -> Integer }) (λthunk@$_{30}$@ : Unit@$_{9}$@ .
n@$_{24}$@)) (λthunk@$_{31}$@ : Unit@$_{9}$@ .
0)) Unit@$_{10}$@))) (λthunk@$_{34}$@ : Unit@$_{9}$@ .
False@$_{2}$@)) Unit@$_{10}$@) (λarg@$_{19}$@ : Integer .
λarg@$_{20}$@ : Integer .
(λb@$_{21}$@ : Bool .
(((ifThenElse { Bool@$_{0}$@ }) b@$_{21}$@) True@$_{1}$@) False@$_{2}$@) ((lessThanEqInteger arg@$_{19}$@) arg@$_{20}$@)) {...})
(λarg_0@$_{36}$@ : Integer .
Λa@$_{37}$@ : *.
λcase_Fixed@$_{38}$@ : Integer -> a@$_{37}$@ .
λcase_Never@$_{39}$@ : a@$_{37}$@ .
case_Fixed@$_{38}$@ arg_0@$_{36}$@)) (Λa@$_{40}$@ : *.
λcase_Fixed@$_{41}$@ : Integer -> a@$_{40}$@ .
λcase_Never@$_{42}$@ : a@$_{40}$@ .
case_Never@$_{42}$@)) (λx@$_{43}$@ : ∀a@$_{44}$@ : *.(Integer -> a@$_{44}$@) -> (a@$_{44}$@ -> a@$_{44}$@) .
x@$_{43}$@) { ∀a@$_{45}$@ : *.a@$_{45}$@ -> a@$_{45}$@ }) (Λa@$_{46}$@ : *.
λcase_Unit@$_{47}$@ : a@$_{46}$@ .
case_Unit@$_{47}$@)) (λx@$_{48}$@ : ∀a@$_{49}$@ : *.a@$_{49}$@ -> a@$_{49}$@ .
x@$_{48}$@)) (λarg@$_{6}$@ : Integer .
λarg@$_{7}$@ : Integer .
(λb@$_{8}$@ : Bool .
(((ifThenElse { Bool@$_{0}$@ }) b@$_{8}$@) True@$_{1}$@) False@$_{2}$@) ((greaterThanEqInteger arg@$_{6}$@) arg@$_{7}$@)) {...})
(Λa@$_{51}$@ : *.
λcase_True@$_{52}$@ : a@$_{51}$@ .
λcase_False@$_{53}$@ : a@$_{51}$@ .
case_True@$_{52}$@)) (Λa@$_{54}$@ : *.
λcase_True@$_{55}$@ : a@$_{54}$@ .
λcase_False@$_{56}$@ : a@$_{54}$@ .
case_False@$_{56}$@)) (λx@$_{57}$@ : ∀a@$_{58}$@ : *.a@$_{58}$@ -> (a@$_{58}$@ -> a@$_{58}$@) .
x@$_{57}$@)
\end{appendixcode}

\end{changemargin}

\fi

\end{document}